\documentclass[usenatbib]{mn2e}

\usepackage{graphicx}
\usepackage{multirow}
\usepackage{natbib}
\usepackage{lineno}
\def\citeN{\citet}
\def\cite{\citep}

\newcommand{\altaffilmark}[1]{$^{#1}$}
\newcommand{\altaffiltext}[2]{$^{#1}$ #2\\}

\footnotesize
\newdimen\digitwidth    %define ! a one digit width for tables
\setbox0=\hbox{\rm0}
\digitwidth=\wd0
\catcode`!=\active
\def!{\kern\digitwidth}
\normalsize
\title[Discovery of pulsars in {\it Fermi} LAT sources]{Discovery of millisecond pulsars in radio searches of southern {\it Fermi} LAT sources}
\author[M.~J.~Keith et al.]
{M.~J.~Keith\altaffilmark{1}\thanks{Email: mkeith@pulsarastronomy.net},
S.~Johnston\altaffilmark{1},
P.~S.~Ray\altaffilmark{2},
E.~C.~Ferrara\altaffilmark{3},
P.~M.~Saz~Parkinson\altaffilmark{4},\newauthor
\"O.~\c{C}elik\altaffilmark{3,5,6},
A.~Belfiore\altaffilmark{7,8,4},
D.~Donato\altaffilmark{5,9},
C.~C.~Cheung\altaffilmark{10},
A.~A.~Abdo\altaffilmark{10},\newauthor
F.~Camilo\altaffilmark{11},
P.~C.~C.~Freire\altaffilmark{12},
L.~Guillemot\altaffilmark{12},
A.~K.~Harding\altaffilmark{3},
M.~Kramer\altaffilmark{13,12},\newauthor
P.~F.~Michelson\altaffilmark{14},
S.~M.~Ransom\altaffilmark{15},
R.~W.~Romani\altaffilmark{14},
D.~A.~Smith\altaffilmark{16},
D.~J.~Thompson\altaffilmark{3},\newauthor
P.~Weltevrede\altaffilmark{13},
K.~S.~Wood\altaffilmark{2},
\\
\altaffiltext{1}{Australia Telescope National Facility, CSIRO, Epping NSW 1710, Australia}
\altaffiltext{2}{Space Science Division, Naval Research Laboratory, Washington, DC 20375, USA}
\altaffiltext{3}{NASA Goddard Space Flight Center, Greenbelt, MD 20771, USA}
\altaffiltext{4}{Santa Cruz Institute for Particle Physics, Department of Physics and Department of Astronomy and Astrophysics, University of California at Santa Cruz, Santa Cruz, CA 95064, USA}
\altaffiltext{5}{Center for Research and Exploration in Space Science and Technology (CRESST) and NASA Goddard Space Flight Center, Greenbelt, MD 20771, USA}
\altaffiltext{6}{Department of Physics and Center for Space Sciences and Technology, University of Maryland Baltimore County, Baltimore, MD 21250, USA}
\altaffiltext{7}{INAF-Istituto di Astrofisica Spaziale e Fisica Cosmica, I-20133 Milano, Italy}
\altaffiltext{8}{Universit\`{a} di Pavia, Dipartimento di Fisica Teorica e Nucleare (DFNT), I-27100 Pavia, Italy}
\altaffiltext{9}{Department of Physics and Department of Astronomy, University of Maryland, College Park, MD 20742, USA}
\altaffiltext{10}{National Research Council Research Associate, National Academy of Sciences, Washington, DC 20001, resident at Naval Research Laboratory, Washington, DC 20375, USA}
\altaffiltext{11}{Columbia Astrophysics Laboratory, Columbia University, New York, NY 10027, USA}
\altaffiltext{12}{Max-Planck-Institut f\"ur Radioastronomie, Auf dem H\"ugel 69, 53121 Bonn, Germany}
\altaffiltext{13}{Jodrell Bank Centre for Astrophysics, School of Physics and Astronomy, The University of Manchester, M13 9PL, UK}
\altaffiltext{14}{W. W. Hansen Experimental Physics Laboratory, Kavli Institute for Particle Astrophysics and Cosmology, Department of Physics and SLAC National Accelerator Laboratory, Stanford University, Stanford, CA 94305, USA}
\altaffiltext{15}{National Radio Astronomy Observatory (NRAO), Charlottesville, VA 22903, USA}
\altaffiltext{16}{Universit\'e Bordeaux 1, CNRS/IN2p3, Centre d'\'Etudes Nucl\'eaires de Bordeaux Gradignan, 33175 Gradignan, France}
}
\date{}
\begin{document}

\maketitle
\newcommand{\setthebls}{
%                 de-comment this line for double spacing:
%\baselineskip=20pt
}

\setthebls

\begin{abstract} 
Using the Parkes radio telescope we have carried out deep observations of eleven unassociated gamma-ray sources.
Periodicity searches of these data have discovered two millisecond pulsars, PSR J1103--5403 (1FGL J1103.9--5355) and PSR J2241--5236 (1FGL J2241.9--5236), and a long period pulsar, PSR J1604--44 (1FGL J1604.7--4443).
In addition we searched for but did not detect any radio pulsations from six gamma-ray pulsars discovered by the {\it Fermi} satellite to a level of $\sim 0.04$ mJy (for pulsars with a 10\% duty cycle).

Timing of the millisecond pulsar PSR J1103--5403 has shown that its position is 9\arcmin\ from the centroid of the gamma-ray source.
Since these observations were carried out, independent evidence has shown that 1FGL J1103.9--5355 is associated with the flat spectrum radio source PKS 1101--536.
It appears certain that the pulsar is not associated with the gamma-ray source, despite the seemingly low probability of a chance detection of a radio millisecond pulsar.
We consider that PSR J1604--44 is a chance discovery of a weak, long period pulsar and is unlikely to be associated with 1FGL J1604.7--4443.
PSR J2241--5236 has a spin period of 2.2~ms and orbits a very low mass companion with a 3.5 hour orbital period.
The relatively high flux density and low dispersion measure of PSR J2241--5236 makes it an excellent candidate for high precision timing experiments.
The gamma-rays of 1FGL J2241.9--5236 have a spectrum that is well modelled by a powerlaw with exponential cutoff, and phase-binning with the radio ephemeris results in a multi-peaked gamma-ray pulse profile.
Observations with Chandra have identified a coincident X-ray source within 0.1\arcsec of the position of the pulsar obtained by radio timing.

\end{abstract}


\section{Introduction}
The Large Area Telescope (LAT, \citealp{fermi_lat}) aboard the {\it Fermi} Gamma-ray Space Telescope has detected more than six hundred unassociated gamma-ray sources amongst the $\sim 1500$ sources in the first year point source catalogue (1FGL, \citealp{1FGL}).
Of the gamma-ray point sources which are associated with known astronomical objects, almost all fall into one of two groups: radio-loud nuclei of distant galaxies and energetic pulsars \cite{agn_catalogue,fermi_pulsar_cat}.
In the last 15 years there has been great interest in observing unassociated gamma-ray sources with the aim of detecting radio pulsars (e.g. \citealp{rhr+02,kjk+08,crr+09,fermiradiofaint,rkp+10}).
Prior to the launch of {\it Fermi}, the positional error-circles of gamma-ray sources have generally been larger than the primary beam of large radio telescopes (at the typical observing frequency of 1400~MHz).
Because of this, these searches have required many pointings to cover each source, limiting the sensitivity of the searches and the number of sources that can be covered.

The increased sensitivity and spatial resolution of the LAT means that the error circles of most sources are comparable to the typical radio beam size, allowing for much deeper radio searches with a single pointing per source.
The radio pulsar community is conducting a coordinated effort to observe the unassociated gamma-ray sources in the radio band.
There are two main benefits for identifying gamma-ray sources as radio pulsars.
Firstly, we can remove these objects from the unassociated gamma-ray list and therefore hope to better understand of the nature of the remaining unassociated sources.
Secondly, in most cases, the radio detections are essential for studying the gamma-ray pulsars themselves.
Radio observations of a pulsar can give a precise ephemeris for phase binning the gamma-ray photons to form a gamma-ray pulse profile.
The phase-resolved intensity and spectra of gamma-ray pulsars are providing a great deal of insight into the pulsar magnetosphere and allowing for distinction between the various models of gamma-ray emission, especially when compared with radio profiles \cite{rw10}.

Additionally, the LAT sensitivity has allowed for the discovery of 24 pulsars by their gamma-ray emission alone \cite{fermi_16bsp,fermi_8bsp}.
Understanding the fraction of gamma-ray pulsars with detectable radio emission has implications for radio and gamma-ray emission models \cite{widebeams}.
Deep observations of these gamma-ray pulsars are therefore required to confirm or rule out pulsed radio emission.

The pulsar population is typically categorised based on the observed spin period, $P$, and its derivative, $\dot P$.
There are two classes of pulsars for which the rate of rotational energy loss, $\dot E \propto \dot P P^{-3}$, is high enough to enable gamma-ray emission.
The most energetic are `young' pulsars, such as the Vela pulsar (PSR J0835--4510), which typically have spin periods between 30 and 300 ms, and period derivatives above $10^{-15}$ s\,s$^{-1}$.
In contrast, millisecond pulsars (MSPs) such as PSR J0218+4232, have very small period derivatives but high $\dot E$ due to their fast rotation periods, often below 10~ms.
Whilst $\dot E$ appears to give a good measure for potential for gamma-ray emission, sensitivity is also limited by the pulsar distance.

This paper presents a search with the Parkes radio telescope for young pulsars and MSPs, targeting unassociated gamma-ray sources and gamma-ray pulsars discovered by the {\it Fermi} telescope.
\begin{table*}
\caption{
\label{selected}
List of the {\it Fermi} sources targeted in this search. Columns are the 1FGL catalogue name, the coordinates of the observation (J2000), the offset from the 1FGL position, the date of observation.
The rotation period and dispersion measure are listed for the three pulsars discovered in this search$^{\dagger}$.
}
{\tt \footnotesize
\begin{tabular}{lllllll}
\hline
1FGL         & RA & Dec             & Offset & Obs Date  & Period & DM \\
             & (h~~m~~s) & (\degr ~~\arcmin ~~\arcsec)    & (\arcmin)&   & (s) & (cm$^{-3}\,$pc) \\
\hline
J0823.3$-$4248 & 08:23:26.6 &$-$42:50:38.4 & ~2.2  & 2009-08-03 & -- \\
J1018.6$-$5856 & 10:18:30.2 &$-$58:57:47.5 & ~2.1  & 2009-04-14 & -- \\
J1045.2$-$5942 & 10:45:11.1 &$-$59:43:10.2 & ~0.9  & 2009-04-15 & -- \\
J1103.9$-$5355 & 11:03:53.8 &$-$53:55:58.8 & ~0.3  & 2009-08-01 & 0.0033927 & 104 \\
J1514.1$-$4945$^{\dagger}$ & 15:14:20.5 &$-$49:44:30.1 & ~2.3  & 2009-04-13 & -- \\
J1536.5$-$4949 & 15:36:26.4 &$-$49:47:48.5 & ~1.9  & 2009-04-15 & -- \\
J1604.7$-$4443 & 16:04:32.9 &$-$44:42:43.2 & ~2.0  & 2009-08-01 & 1.3892    & 175 \\
J1620.8$-$4928c& 16:21:05.5 &$-$49:30:32.4 & ~3.0  & 2009-08-03 & -- \\
J1636.4$-$4737c& 16:35:31.5 &$-$47:28:55.2 & 13.0  & 2009-04-13 & -- \\
J1640.8$-$4634c& 16:39:59.0 &$-$46:44:24.0 & 13.0  & 2009-08-02 & -- \\
J2241.9$-$5236 & 22:41:55.2 &$-$52:36:39.6 & ~0.3  & 2009-08-01 & 0.0021866 & 11  \\
\hline
\end{tabular}
}
\\$^{\dagger}$A pulsar was discovered in 1FGL~J1514.1$-$4945 in another
search at Parkes, as discussed in Kerr et al. (in prep).
\end{table*}

\section{Source selection}
In March of 2009 we selected sources from a preliminary list of {\it Fermi} point sources for which no formal association with a known astronomical object had been made.
To best suit observation with the Parkes radio telescope, we selected sources with a declination south of $-40^\circ$ and an error circle with diameter smaller than the 14 arcminute beam width of Parkes at 1400~MHz.
These criteria allowed us to select 11 sources, as shown in Table \ref{selected}.
As the {\it Fermi} LAT all-sky survey continues, gamma-ray photon statistics increase and localisation improves.
Therefore the positions observed in this search are offset somewhat from the more accurate position as published the 1FGL catalogue and these offsets are shown in the table.
Additionally, we chose to observe six southern gamma-ray pulsars that had not been deeply searched using radio telescopes, with the aim of detecting any radio pulsations.
The observations of these pulsars are detailed in Table \ref{blindsearch}.
\section{Observation and data analysis}

Each of the selected sources was observed using the Parkes 64~m antenna and the centre beam of the 21-cm `Multibeam' receiver, which has a noise equivalent flux density of $\sim$40~Jy.
The BPSR digital filterbank system was used which has an observing bandwidth of 340~MHz centred at 1352~MHz and split into 870 channels.
Each channel is sampled with 2-bits of precision every 64~$\mu$s.
In order that we might place stringent limits on the detectability of radio pulses from these objects, we observed for 4.8 hours per target.
We determine the minimum detectable flux density according to the standard interpretation of the radiometer equation,
\begin{equation}
S_{\rm min} = \frac{\sigma (T_{\rm sys} + T_{\rm sky})}{G \sqrt{2 B \tau_{\rm obs}}} \sqrt{\frac{W_{\rm frac}}{1-W_{\rm frac}}}.
\label{smin}
\end{equation}
For this paper we assume: system temperature, $T_{\rm sys} = 23$ K; sky temperature $T_{\rm sky}=2.5$ K; gain, $G = 0.735$; bandwidth $B=340$ MHz; observing time $\tau_{\rm obs} = 17200$ s; effective pulse duty cycle, $W_{frac}=0.1$; and the signal-to-noise threshold, $\sigma = 8$.
For a pulsar close to the centre of the beam, this gives a detection threshold of $\sim0.03$~mJy.
A detailed discussion of the sensitivity of the BPSR system can be found in \citeN{hitrun_1}.

For the known gamma-ray pulsars we folded the data using a rotational ephemeris provided by the {\it Fermi} collaboration~\cite{rkp+10}, searching in dispersion measure (DM) from 0 to 1000~cm$^{-3}\,$pc.
The data taken on unassociated 1FGL sources were searched for periodic signals using the {\sc hitrun} pipeline as described in \citeN{hitrun_1}.
Again, DMs up to 1000~cm$^{-3}\,$pc were searched.
We have searched the data attempting to correct for the period drifts due to acceleration of the pulsar if in a binary system.
The acceleration searching was performed using {\sc presto}, using the algorithm described in \citeN{rem02}.
The frequency drift range searched was $\pm150$ Fourier bins, corresponding to an acceleration range of $\pm0.3$~${\rm m}\,{\rm s}^{-2}$ for a pulsar with a rotational frequency of 500~Hz\footnote{The acceleration range goes as 1/$\nu$}.
Additionally we have done the same processing on the first 2100~s of each observation, giving an acceleration range of $\pm20.4$~${\rm m}\,{\rm s}^{-2}$ for the same pulsar but a reduced sensitivity of 0.1~mJy.
Further processing to remove the effects of binary motion will likely be done in the future.
From the candidates presented by the software, three were selected for re-observation at Parkes, and have been confirmed to be radio pulsars\footnote{A fourth
pulsar, detectable in our observation of 1FGL~J1514.1$-$4945, was
discovered in the search described by Kerr et al. (in prep).}.

\subsection{Simulations}
\label{sim}
Na\"ively, one might assume that there is little chance of detecting rare objects such as MSPs in so few pointings, unless the MSP is indeed the source of the gamma rays.
In order to determine the probability that any of our detections could be unrelated to the gamma-ray sources, we chose to simulate the Galactic pulsar population with the {\sc psrpop} software, using the method described in \citeN{hitrun_1}, based on the model of \citeN{lfl+06}.
For the non-MSP population our model has a log-normal spin period distribution (with mean and variance in log-space of $2.71$ and $0.34$ respectively), a power-law luminosity distribution (with index $-0.59$, cutting off below 0.1~mJy~kpc$^2$), an exponential distribution for the height above the Galactic plane (with scale 0.33~kpc) and using the radial distribution given by \citeN{yk04}.
For the MSPs, we used a similar model except that, to better match the known MSP population, the scale height was increased to 0.5~kpc and the period distribution was empirically matched to the population of known MSPs.
The total number of pulsars simulated was limited such that the number of pulsars discovered in a simulated Parkes Multibeam Pulsar Survey \cite{mlc+01} matched the real discovery count.

We then simulated the 11, 4.8-hour pointings of our search that did not point at known gamma-ray pulsars, and recorded the number of pulsars discovered.
This was repeated 1000 times with new simulations of the Galactic pulsar population each time, based on different random seeds.
The mean detection rates were 1.3 non-MSPs and 0.05 MSPs in 11 pointings.
Therefore it is likely that we could discover at least one non-MSP in our searches, and the discovery of an unassociated MSP is certainly not ruled out.

\section{Results and discussion}
None of the six pulsars discovered by their pulsed gamma-rays, as listed in Table \ref{blindsearch}, was detected in our radio searches.
The discovery of these pulsars and searches for their radio pulsations are detailed in \citeN{fermi_16bsp} and \citeN{fermi_8bsp}.
Analysis of the other 11 observations has yielded the discovery of three radio pulsars, and the detection of the MSP J1514--49 (discussed in Kerr et al., in prep).
Table \ref{selected} presents the spin period and DM of the pulsars discovered.
PSRs J1103-5403 and J2241--5236 have spin periods less than 5~ms and therefore belong to the class of MSPs.
Although we initially expected to discover mainly young, energetic pulsars in this search, we are finding that many MSPs are being detected in gamma rays \cite{fermi_msps}.
PSR J1604--44 has a spin period of 1.4~s, typical of old, but not recycled, pulsars which are not commonly associated with gamma-ray emission.
In this section we discuss the non-detections, and present relevant details of the gamma-ray and radio analysis of PSRs J1103--5403, J1604--44 and J2241--5236.

\begin{table}
\caption{
\label{blindsearch}
The six gamma-ray selected pulsars that were targeted for radio observations as part of this work.
}
{\tt \footnotesize
\begin{tabular}{llll}
\hline
PSR & RA& Dec & Obs Date\\
 & (h~~~m~~~s) & (\degr~~~\arcmin~~~\arcsec) & \\
\hline
J1023$-$5746 & 10:23:10.6 & $-$57:44:20.0 & 2009-04-13\\
J1044$-$5737 & 10:44:33.3 & $-$57:37:15.0 & 2009-08-02\\
J1413$-$6205 & 14:13:14.2 & $-$62:04:33.6 & 2009-08-02\\
J1429$-$5911 & 14:30:02.2 & $-$59:11:20.4 & 2009-08-03\\
J1732$-$31   & 17:32:40.4 & $-$31:36:35.3 & 2009-04-14\\
J1809$-$2332 & 18:09:50.2 & $-$23:32:23.0 & 2009-04-15\\
\hline
\end{tabular}
}
\end{table}
\subsection{Non-detections}
No radio pulsars were detected in observations of seven of the unassociated point sources and all six of the gamma-ray pulsars.
1FGL J1636.4$-$4737c and J1640.8$-$4634c are confused with diffuse Galactic emission and are now considered to lie a considerable distance from the observed location, therefore our radio observations are not constraining.
For the 11 other sources we are confident that any radio pulsations associated with the source must be either be below our minimum detectable flux density, hidden by effects of the interstellar medium (scattering or scintillation), or by the effects of binary motion.
For the sources for which gamma-ray pulses have not been detected, our Fourier based search method is only maximally sensitive to solitary pulsars, or those with orbital periods of greater than a few days.
Although our search has used some acceleration compensation, for pulsars with large changes in the apparent pulse period our sensitivity will be considerably reduced.

\subsection{PSR J1103--5403 and 1FGL J1103.9--5355}
\subsubsection{PSR J1103--5403}
\begin{figure}
\includegraphics[height=8cm]{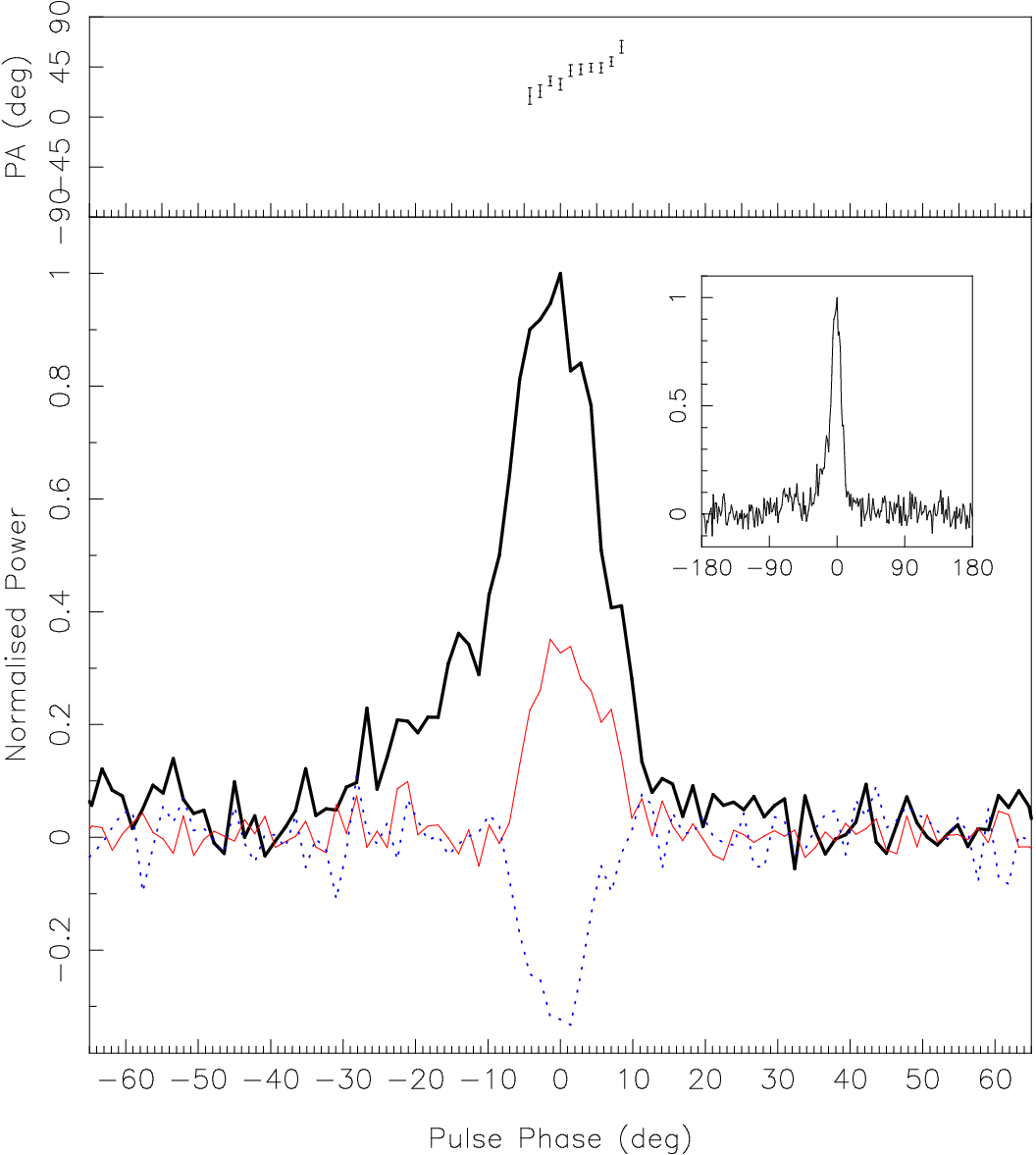}
\caption{
\label{1103_profile}
The 1369~MHz integrated pulse profile of PSR J1103--5403 clipped to a region round the pulse peak.
Total intensity is shown as a thick line, with linear polarisation as a thin solid line and circular polarisation as a dotted line.
Flux density is measured as a fraction of the pulse peak.
The polarisation position angle variation over the pulse is shown in the upper panel.
The polarisation position angles shown are not absolute.
The inset figure shows the total intensity profile over the entire 360\degr.
}
\end{figure}

PSR J1103--5403 is a solitary pulsar with spin period of 3.4~ms and DM 104 cm$^{-3}\,$pc.
Timing parameters are given in Table \ref{1103_table}.
These parameters indicate that PSR J1103--5403 is a MSP at a distance of 2.5~kpc, using the electron density model of \citeN{cl02}.
The average profile appears relatively simple, with a small amount of linear and circular polarisation, as shown in Figure \ref{1103_profile}.

Intriguingly, since its selection for this survey, the gamma-ray source 1FGL J1103.9$-$5355 has been associated with the flat spectrum radio source PKS 1101--536 \cite{msm+10}, believed to be an active galactic nucleus (AGN).
Since both MSPs and AGNs are strong candidates for gamma-ray emission, further analysis is required to determine which is the true source of 1FGL J1103.9$-$5355.
Since the discovery, our timing observations constrain the position of the pulsar to 11h03m33s $-54$\degr03\arcmin43\arcsec, well outside the 2.4\arcmin\ error radius (95\% confidence) of 1FGL J1103.9--5355.
Nevertheless, as described below, we have performed a careful analysis of the gamma-ray data to confirm that the sources are distinct.

\begin{table}
\caption{\label{1103_table}
Astrometric and rotational parameters from the timing of PSR J1103--5403.
Values in parentheses are the nominal 1$\sigma$ uncertainties in the least significant digits quoted.
}
\begin{tabular}{ll}
\hline\hline
\multicolumn{2}{c}{PSR J1103--5403} \\
\hline
Right ascension, $\alpha$\dotfill &  11:03:33.294(3) \\
Declination, $\delta$\dotfill & $-$54:03:43.245(16) \\
Dispersion measure, $DM$ (cm$^{-3}$pc)\dotfill & 103.915(3) \\
Rotation measure, $RM$ (rad m$^{-2}$)\dotfill & $-103$(21) \\
Pulse frequency, $\nu$ (s$^{-1}$)\dotfill & 294.749654528(3) \\
First derivative of pulse frequency, $\dot{\nu}$ (s$^{-2}$)\dotfill & $-$3.2(8)$\times 10^{-16}$ \\
Epoch of model (MJD)\dotfill & 55300 \\
\\
Validity of solution (MJD) \dotfill & 55044.1---55459.9 \\
Rms of residuals ($\mu s$)\dotfill & 6.42 \\
Clock correction procedure\dotfill & TT(TAI) \\
Solar system ephemeris model\dotfill & DE405 \\
Coordinate time standard\dotfill & TCB \\
\\
Flux density at 1.4~GHz ($S_{1400}$, mJy)\dotfill & 0.18(4) \\
Width at 50$\%$ of pulse peak ($W_{50}$, $\mu$s)\dotfill & 140 \\
\\
Distance$^*$ (kpc)\dotfill & 2.5 \\
Characteristic age (yr) \dotfill & $1.4 \times 10^{10}$ \\
Surface magnetic field strength (G) \dotfill & $1.1 \times 10^{8}$ \\
Rate of energy loss, $\dot E$ (erg$\,{\rm s}^{-1}$)\dotfill & $3.8\times10^{33}$\\
\hline
\end{tabular}
$^*$ assumes the \citeN{cl02} model of electron density.

\end{table}

\subsubsection{Gamma-ray data selection\label{data_selection}}

\begin{figure}
\centering
\includegraphics[clip=true,trim=1.5cm 3.2cm 0 3.2cm,width=8.0cm,angle=-90]{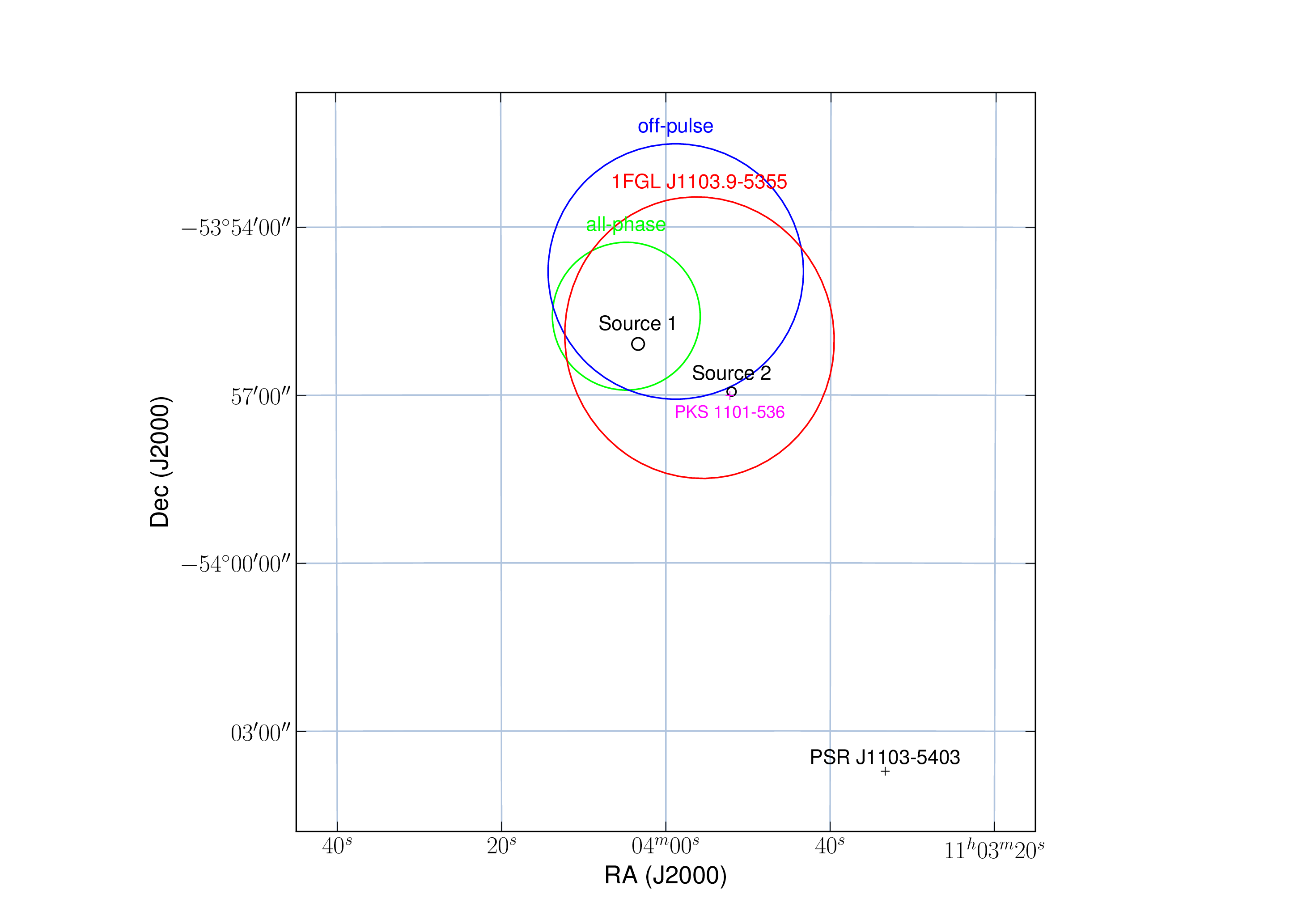}\\
\caption{
%{\it Fermi} LAT counts map of the region around 1FGL\,J1103.9-5355 in the 0.1-300 GeV energy range. The two {\it Swift} sources close to the gamma-ray source are indicated by small white circles and labelled 1 and 2. The red, blue, and green circles show the 95\% (statistical) errors around the position from the 1FGL catalogue, `off-pulse', and `all-phase' analyses respectively. The cyan cross shows the position of the known active radio galaxy PKS\,1101--536, which is coincident with {\it Swift} source number 2. The brown line shows the timing position of PSR\,J1103-5403.
The measurements of position of 1FGL\,J1103.9-5355 and nearby Swift sources shown with the radio position of PKS\,1101--536 and PSR\,J1103-5403.
The large circles show the 95\% confidence limit of the position from the 1FGL catalogue, `off-pulse', and `all-phase' analyses as labelled. 
The two {\it Swift} sources close to the gamma-ray source are indicated by small circles and labelled Source 1 and 2.
The cross labelled PKS\,1101--536 shows the position of the known active radio galaxy, and is coincident with {\it Swift} source number 2.
The cross labelled PSR\,J1103-5403 marks the position of the pulsar obtained from radio timing.
}
\label{localization}
\end{figure}

We used 21 months of {\it Fermi} LAT data taken in sky-scanning survey mode between 4 August
2008 -- 4 May 2010 (MET\footnote{Mission Elapsed Time, or seconds since 00:00 UTC on 2001 January 1} 239557417-- 294680617). This extends the data set used for the 1FGL catalogue by an additional ten months.
As in the 1FGL catalogue, we used only pass 6v3 {\it diffuse} class photons. We selected events within a 10$^\circ$ radius of 1FGL\,J1103.9--5355 in the energy range (100 MeV--300 GeV).
The software used for the analysis was the LAT Science Tools v9r15p2, available from the {\it Fermi} Science Support Centre (FSSC)
website\footnote{http://fermi.gsfc.nasa.gov/ssc/}, with the instrument response functions PASS6\_V3.

The pulsar PSR\,J1057--5226~\citep{3EGRETPulsars} is located 1.8$^\circ$ from 1FGL\,J1103.9--5355 and its large gamma-ray flux could therefore have an effect on our spectral and localisation results. In order to remove (or at least reduce) the effect of the pulsar, we selected the subset of events falling in the off-pulse region of PSR\,J1057--5226, according to an updated ephemeris for the pulsar derived from LAT data covering the entire 21-month period of observations~\footnote{The timing model used in this study will be made available on the servers of the FSSC.}. We refer to this reduced data set, consisting of $\sim$60\% of the events, as the `off-pulse' data set. When the entire data set is used, we will refer to this as the `all-phase' data set.

\subsubsection{Gamma-ray spectrum and localisation}
\label{Sec:J1103Spect}
We performed an unbinned likelihood analysis using the LAT Science Tool {\tt gtlike}, adopting the standard
models for the diffuse emission components, both Galactic ({\tt gll\_iem\_v02.fit}) and isotropic  
({\tt isotropic\_iem\_v02.txt}). All 1FGL sources within 15$^\circ$ of 1FGL\,J1103.9--5355 were included in the
spectral model. We first assumed a power-law spectrum for all sources and then refined the model by
introducing an exponential cutoff for 12 of them, after determining, iteratively, by means of a likelihood ratio test, that
the addition of such a parameter resulted in a significant improvement of the model. 
For all sources within 6$^\circ$ of 1FGL\,J1103.9--5355 we fitted the normalisation, while
for PSR J1057--5226 and 1FGL\,J1117.0--5339 all the spectral parameters were allowed to be free.

The addition of a cutoff in the spectrum of 1FGL\,J1103.9--5355 did not lead to a statistically significant increase in the likelihood of the model, and we
therefore modelled the source with a simple power law. Using the `all-phase' sample, we obtained a best fit value for the spectral index of $2.20\pm0.06$, broadly in
agreement with what was reported in the 1FGL catalogue ($2.05\pm0.06$). The resulting (0.1-300 GeV) photon flux is $(6.3\pm0.8)\times10^{-8}$ cm$^{-2}\,$s$^{-1}$.
We repeated our analysis using the `off-pulse' data set and obtained a spectral index of $2.20\pm0.07$ and a photon flux of $(6.2\pm0.8)\times10^{-8}$ cm$^{-2}\,$s$^{-1}$,
consistent with the previous results.

Using the LAT Science Tool {\tt gtfindsrc} we attempted to improve on the 1FGL localisation by performing a maximum likelihood fit on our longer data set and more
refined spectral model. We allowed the parameters of sources within 2$^\circ$ of 1FGL\,J1103.9--5355 to vary, while fixing all the
other parameters to the values obtained from our spectral analysis. In order to check the reliability of this technique, we used {\tt gtfindsrc} and our spectral
model for the region to derive a position for PSR\,J1057--5226. The resulting best fit position is 10h57.96m $-52$\degr26.8\arcmin, with a 95\% (statistical) uncertainty
radius of r95\%=0.84\arcmin, less than 0.18\arcmin away from its well-determined radio position.

For 1FGL\,J1103.9--5355, using the `all-phase' data set, we obtain a best fit position of 11h04.08m $-53$\degr55.6\arcmin, and r95\%=1.3\arcmin. To check for
possible effects from the nearby pulsar PSR\,J1057--5226 (described in Section~\ref{data_selection}) we ran {\tt gtfindsrc} on the
`off-pulse' data set and model, obtaining a best fit position of 11h03.98m $-53$\degr54.8\arcmin, with r95\%=2.3\arcmin, consistent with our previous result. 

Using a $\sim$4 ks 0.2--10 keV {\it Swift} XRT observation (ObsID 00031616001), we searched for possible X-ray counterparts of 1FGL\,J1103.9--5355.
Five sources are detected in the {\it Swift} XRT~\citep{xrt} field of view, but only two are close enough to be consistent with the LAT gamma-ray source (labelled 1 and 2 in Figure~\ref{localization}). Source 1 is located at 11h04m03.37s $-$53\degr56\arcmin05.50\arcsec\ (with r90\%=6.65\arcsec) and has a count rate of $2.63\pm0.99\times10^3$ cts s$^{-1}$, while source 2 is located at 11h03m52.02s
$-53$\degr56\arcmin56.61\arcsec\ (with r90\%=4.76\arcsec) and has a count rate of $13.9\pm2.1\times10^3$ cts$\,{\rm s}^{-1}$. Consistent with the position of
{\it Swift} X-ray source 2 is a bright radio source, PKS\,1101--536, located at 11h03m52.221639(13)s $-53$\degr57\arcmin00.69607(20)\arcsec~\citep{pks1101,msm+10}.
These gamma-ray, X-ray and radio positions are summarised in Figure~\ref{localization}.

\subsubsection{Variability analysis}
\begin{figure}
\centering
\includegraphics[width=1.0\columnwidth,trim=0 5 0 10]{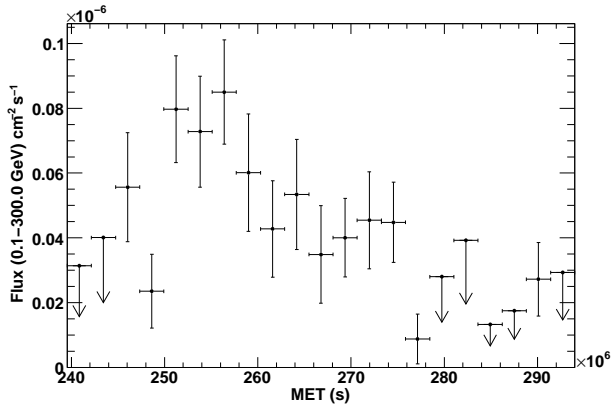}\\
\caption{{\it Fermi} LAT monthly light curve of 1FGL\,J1103.9--5355 in the 0.1-300 GeV energy range.}
\label{lightcurve}
\end{figure}

Since gamma-ray pulsars, unlike AGNs, are typically not variable on long time-scales we investigate the variability characteristics of 1FGL\,J1103.9--5355.
%The 1FGL catalogue implies that this has a probability of $<$1\% of being a steady source (the variability index, V=48.8; \citealp{1FGL}).
The 1FGL catalogue gives a variability index of V=48.8 for this source, implying that it has a probability of $<$1\% of being a steady
source \cite{1FGL}.
Here, we extend the variability analysis to our 21-month data set. 
For this variability analysis, we used the `off-pulse' data set as defined in Section \ref{data_selection} to avoid any variability induced due to the fluctuations in the spectral fit of the nearby strong pulsar PSR J1057--5226.
We obtained a monthly light curve for 1FGL J1103.9--5355 dividing the data set into 30-day intervals and performing a spectral analysis in each time bin.
In each time bin, we used the same spectral model as described in Section \ref{Sec:J1103Spect} for nearby sources and diffuse components, allowing only their normalisation parameters to be free in the fit, and fixing their spectral indices to the best fit value obtained from the analysis of full time period.
We obtained the flux of 1FGL J1103.9--5355 for each time bin it was detected with a test statistic, TS, greater than 9 \cite{mbc+96}, and we calculated a 95\% confidence level upper limit for those time bins that the source was not detected. 
We found that 1FGL J1103.9--5355 is clearly variable as seen from Figure 3; it exhibits an excess emission over its average flux level early in the observation period, while it drops below the detection limit in the first two and five of the last 6 months.
Since the monthly light curve turns into upper limits in some time bins, we repeated the spectral analysis splitting the data set into 4 segments, each  ~160 days, sufficiently long to detect the source in every time bin.
We got the following values for the photon flux of 1FGL J1103.9--5355 (in units of $10^{-8}$cm$^{-2}\,$s$^{-1}$): $9.2\pm1.6$, $10.4\pm1.5$, $4.2\pm1.2$, and $1.1\pm0.7$. The corresponding fit to a constant has a $\chi^{2}=45.65$ with three degrees of freedom, allowing us to reject the hypothesis of a steady
source with a confidence level of $>6\sigma$. A very conservative estimate of the systematics as 20\% of the measured flux still confirms the variability of the source at a 3$\sigma$ level.

\subsubsection{Summary}
The spectral, positional and temporal properties of the gamma-ray source 1FGL J1103.9$-$5355 are inconsistent with originating from PSR J1103--5403.
It seems most probable that 1FGL J1103.9$-$5355 is associated with PKS 1101--536 as described in \citeN{msm+10}.
Therefore we conclude that the discovery of PSR J1103--5403 was entirely coincidental, despite the seemingly low probability.

\subsection{PSR J1604--44}
PSR J1604--44 has a spin period of 1.4~s, which is not typical of the pulsars detected by {\it Fermi}.
Since its selection for this survey, 1FGL J1604.7--4443 has been associated with the flat spectrum radio source PMN J1604--4441 \cite{agn_catalogue}.
As discussed in Section \ref{sim}, the long integration time of our survey means that it is possible for us to detect one or more pulsars unrelated to the gamma-ray sources.
Therefore we believe it is quite likely that this is a coincidental Galactic plane pulsar along the same line of sight as 1FGL J1604.7--4443 and PMN J1604--4441.
Unfortunately the flux density of this pulsar is so low that the normal timing procedure is impractical with the Parkes radio telescope, and so we are unable to provide further details here.

\subsection{PSR J2241--5236}
\subsubsection{Radio timing}
The pulsar J2241--5236 has a spin period of 2.19~ms and a period derivative of $6.6 \times 10^{-21}$ s\,s$^{-1}$.
This places the pulsar firmly amongst the `recycled pulsars', which have undergone a period of accretion-induced spin-up from a companion star.
A full timing solution, and some standard derived parameters for J2241--5236 are given in Table \ref{2241_table}.
The energy loss rate, $\dot E$ is $2.5 \times 10^{34}$ erg$\,{\rm s}^{-1}$, well within the typical values for gamma-ray emitting pulsars.
An oft-used gamma-ray detectability measure is $\dot E^{1/2} /d^2$, which if we accept the DM-derived distance of 0.5~kpc, is $6 \times 10^{17}$~erg$^{1/2}\,$pc$^{-2}\,{\rm s}^{-1/2}$, typical of the known gamma-ray MSPs \cite{fermi_pulsar_cat}.

\begin{figure}
\includegraphics[height=8cm]{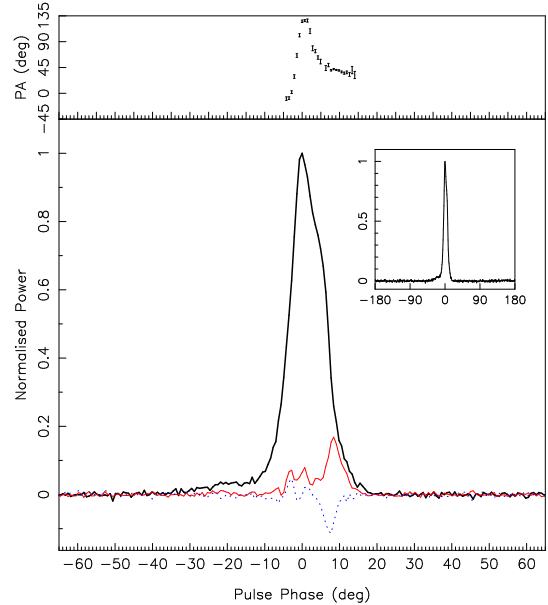}
\caption{
\label{profile}
The 1369~MHz integrated pulse profile of PSR J2241--5236 clipped to a region round the pulse peak.
Total intensity is shown as a thick line, with linear polarisation as a thin solid line and circular polarisation as a dotted line.
Flux density is measured as a fraction of the pulse peak.
The polarisation position angle variation over the pulse is shown in the upper panel.
The polarisation position angles shown are not absolute.
The inset figure shows the total intensity profile over the entire 360\degr.
}
\end{figure}

Figure \ref{profile} shows the polarisation profile of PSR J2241--5236, averaged over many hours of observation.
The profile appears to be composed of several overlapping components.
The pulse width at $50\%$ of the peak is 68~$\rm \mu$s and the profile shows some linear and circular polarisation, strongest towards the trailing edge of the pulse.

The pulsar is in a circular orbit with an orbital period of 3.5 hours.
The mass function is $8.6\times 10^{-7}$~M$_{\sun}$, which assuming a pulsar mass of 1.35~M$_{\sun}$, implies a minimum companion mass of 0.012~M$_{\sun}$.
Excluding planetary objects, this is the smallest mass function seen in pulsars outside of globular clusters.
The standard explanation for the companion would be a white dwarf, from which mass has been stripped by the pulsar wind~\cite{krst88,pebk88,vv88}.
Similar to the PSR J0610--2100 \cite{bjd+06}, PSR J2241--5236 does not show any eclipse or DM variability throughout its orbit.
It is likely therefore that the orbit is sufficiently inclined to the line of sight such that we do not `see' the material removed from the companion.
For this reason, we deem it unlikely that any Shapiro delay will be measured in this system.

Given the high flux density and low DM of PSR J2241--5236 we have sought to determine the prospects for measuring proper motion and parallax effects, which would provide a model-independent distance to the pulsar.
To do this we have simulated a 4 year timing campaign, observing every 30 days, with 0.7$\mu$s RMS in the residuals.
The real RMS in such a campaign will likely be dominated by unmodelled `timing noise', however this simulation does show the potential for future timing of PSR J2241--5236.
For our simulation we assumed the DM distance of 0.5~kpc and a velocity of 100~km s$^{-1}$.
The simulated data were fit for frequency, frequency derivative, position, binary parameters, proper motion and parallax.
The error in the proper motion was less than 0.1~mas yr$^{-1}$, equivalent to a transverse velocity error of less than 1~km s$^{-1}$.
The parallax measurement gave an error of 0.4~mas, which equates to a distance error of 80~pc.
Although real observations may well be adversely affected by additional sources of noise, this simple simulation indicates that PSR J2241--5236 has the potential to be studied in great detail.
We intend to continue monitoring this pulsar for several years to measure the parameters such as parallax, as well as to investigate the prospects of using PSR J2241--5236 in a pulsar timing array \cite{haa+10}.

\begin{table}
\caption{\label{2241_table}
Astrometric, rotational and orbital parameters from the timing of PSR J2241--5236.
Values in parentheses are the nominal 1$\sigma$ uncertainties in the least significant digits quoted.
}
\begin{tabular}{ll}
\hline\hline
\multicolumn{2}{c}{PSR J2241$-$5236} \\
\hline
Right ascension, $\alpha$\dotfill &  22:41:42.01850(5) \\
Declination, $\delta$\dotfill & $-$52:36:36.2260(3) \\
Dispersion measure, $DM$ (cm$^{-3}$pc)\dotfill & 11.41085(3) \\
Rotation measure, $RM$ (rad m$^{-2}$)\dotfill & 14(6) \\
Pulse frequency, $\nu$ (s$^{-1}$)\dotfill & 457.3101497559(3) \\
First derivative of pulse frequency, $\dot{\nu}$ (s$^{-2}$)\dotfill & $-$1.388(10)$\times 10^{-15}$ \\
Epoch of model (MJD)\dotfill & 55044.2 \\
\\
Binary model\dotfill & BT$^\dagger$ \\
Orbital period, $P_b$ (d)\dotfill & 0.1456722395(3) \\
Epoch of periastron, $T_0$ (MJD)\dotfill & 55044.1580909(6) \\
Projected semi-major axis of orbit, $x$ (lt-s)\dotfill & 0.02579537(14) \\
Eccentricity\dotfill & $<1\times10^{-5}$ \\
\\
Validity of solution (MJD)\dotfill & 55044.5---55509.5 \\ 
RMS of residuals ($\mu$s)\dotfill & 0.7\\
Clock correction procedure\dotfill & TT(TAI) \\
Solar system ephemeris model\dotfill & DE405 \\
Coordinate time standard\dotfill & TCB \\
\\
Flux density at 1.4~GHz ($S_{1400}$, mJy)\dotfill & 4.1(1) \\
Width at $50\%$ of pulse peak ($W_{50}$, $\mu$s)\dotfill & 70 \\
Width at $10\%$ of pulse peak ($W_{10}$, $\mu$s)\dotfill & 130 \\
\\
Distance$^*$ (kpc)\dotfill & 0.5 \\
Characteristic age (yr)\dotfill & $5\times10^{9}$\\
Surface magnetic field strength, $B$ (G)\dotfill & $1.2\times10^{8}$\\
Rate of energy loss, $\dot E$ (erg$\,{\rm s}^{-1}$)\dotfill & $2.5\times10^{34}$\\
Mass function ($M_{\sun}$) \dotfill & $8.6844(4)\times 10^{-7}$\\
Minimum companion mass ($M_{\sun}$)\dotfill & 0.012 \\
\hline
\end{tabular}
$^*$ assumes the \citeN{cl02} model of electron density.
$^\dagger$ \citeN{bt76}.
%Note: Figures in parentheses are  the nominal 1$\sigma$ \textsc{tempo2} uncertainties in the least-significant digits quoted.

\end{table}

\subsubsection{X-ray observations}

PSR J2241--5236 was covered by a programme of observing unassociated Fermi-LAT sources with the
Chandra X-ray Observatory. On 2009 August 30, we observed this source for 20 ks with the ACIS-I instrument. The brightest X-ray source within the 1FGL error circle is at (J2000.0) R.A. = 22:41:42.01, Decl. = -52:36:36.2 ($90\%$ radius = 0.3", statistical only),
less than 0.1 arcseconds from the position obtained by radio timing. Therefore we are confident that PSR J2241--5236 is the source of the observed X-rays.

We measured 86 counts (0.5 -- 7 keV) arriving within 5\arcsec of the pulsar position, whereas only 1.75 counts are expected due to background.
Therefore we model the spectrum using Cash statistics without subtracting the background.
The X-ray spectrum was initially fit with an absorbed
single black-body model with temperature, $kT$ = 0.26 $\pm$ 0.04 keV (c-stat/d.o.f. = 174/254),
but clearly left residual emission at $<$0.8 keV unmodelled.
Adding a second black-body ($kT$ = 0.07 $\pm$ 0.04 keV) improved the fit (c-stat/d.o.f. = 167/252)
and did not affect the temperature of the first black-body ($kT$ = 0.31 $\pm$ 0.08 keV). The total $0.5-7$
keV X-ray flux in the latter model was (3.1$^{+0.4}_{-1.6}$) $\times$ 10$^{-14}$ erg cm$^{-2}$ s$^{-1}$. The
effect of Galactic absorption ($N_{\rm H}=1.21 \times 10^{20}$ cm$^{-2}$; \citealp{kbh+05}) in the fits
was negligible, and the unabsorbed flux was 3.3 $\times$ 10$^{-14}$ erg cm$^{-2}$ s$^{-1}$.
Assuming a distance of 0.5 kpc equates to a luminosity of $\sim$1 $\times$ 10$^{30}$ erg s$^{-1}$.
This is less than 1$\%$ of the rotational energy loss rate, and so can be easily powered by the available
spin down energy budget.

\subsubsection{Gamma-ray pulsations}
\begin{figure}
\centering
\includegraphics[width=1.0\columnwidth]{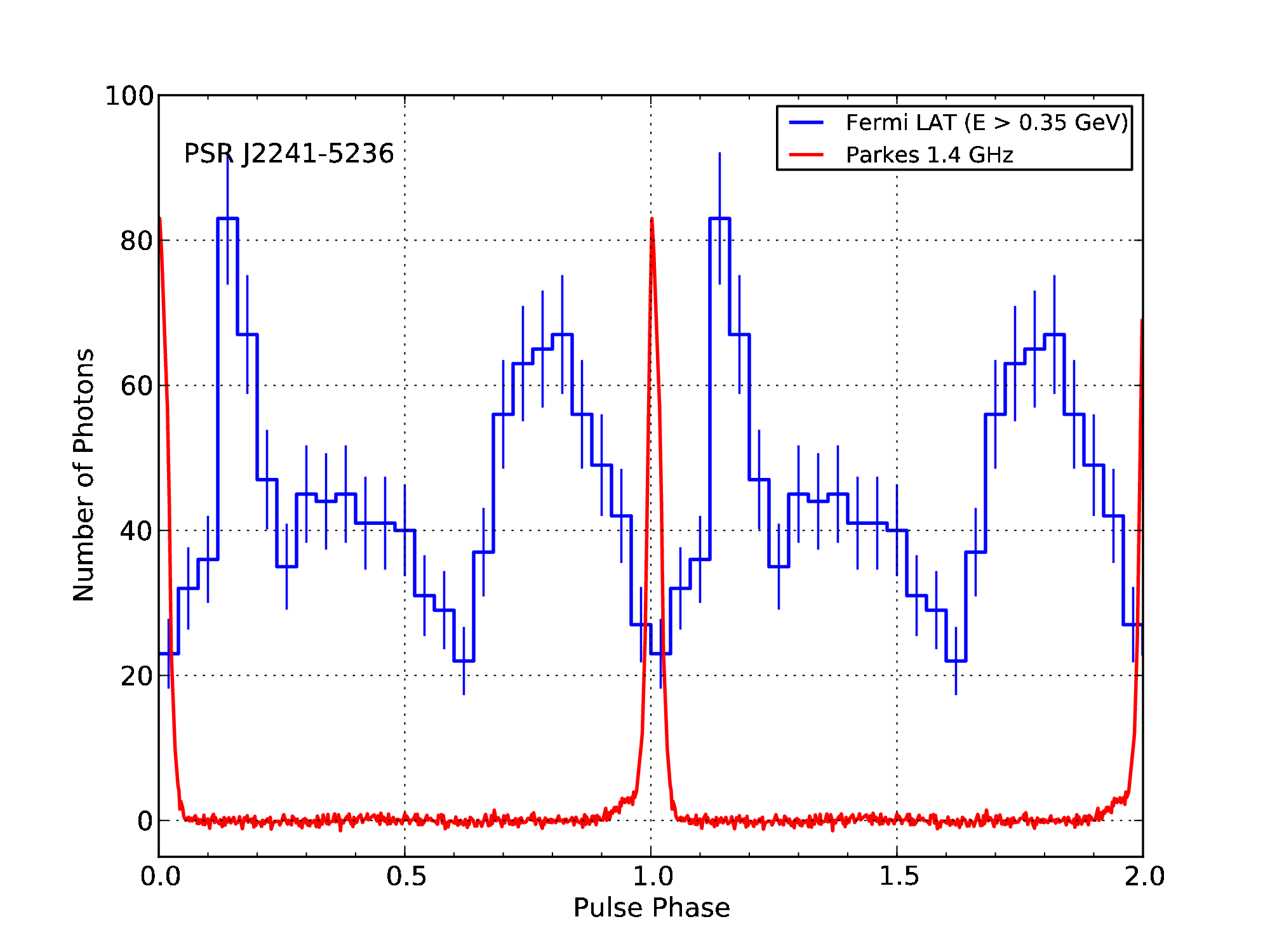}\\
\caption{
Phase aligned radio (solid line) and gamma-ray (histogram with errors) integrated profiles of PSR J2241--5236.
The gamma-rays from 1FGL J2241.9--5236 were phase binned using the ephemeris obtained by radio timing.
Two full periods are shown for clarity.
\label{J2241prof}}
\end{figure}

To search for gamma-ray pulsations, we have phase-binned the gamma-rays from 1FGL J2241.9--5236 using the ephemeris obtained from the radio timing.
To maximise the detection significance we searched over 20 values of radius and 20 values of low energy cut-off.
With an extraction radius of 1.2$^\circ$ and a low energy cut-off of $E>350$ MeV we obtain a H-test parameter of 108, equivalent to a detection significance above 8$\sigma$ \cite{db10}.
The best profile is shown in Figure \ref{J2241prof}, produced from the 1123 photons arriving between MJD 54684 and 55511 that matched our cuts.
We are therefore confident that PSR J2241--5236 is the source of gamma-rays in 1FGL J2241.9--5236.
It is not clear if the gamma-ray profile has two or three components, however we can measure the radio lag (i.e. the separation of the radio peak and the first gamma-ray peak) to be $0.15 \pm 0.01$ in units of pulse phase.

\subsubsection{Gamma-ray spectral analysis}
We analysed the gamma-ray spectrum of 1FGL J2241.9--5236 using the same 21 months time interval from 2008 August 4 to 2010 May 4, as the 1FGL J1103.9--5355 data set and we used the same analysis cuts and methods described in Section~\ref{Sec:J1103Spect}. We analysed a region of 10$^\circ$\ radius around the radio position of the PSR J2241--5236, and modelled all of the 22 1FGL sources within 15$^\circ$\ of the pulsar along with Galactic and isotropic diffuse emission as detailed in Section~\ref{Sec:J1103Spect}. There are no known pulsars in this region. Thus we did not use any cuts on the phase, and we modelled all sources by power law spectral shapes. The spectral parameters for sources within the 10$^\circ$\ of 1FGL J2241.9--5236 were left free in the likelihood fit, and the spectral parameters for other sources were fixed to the values obtained from 1FGL catalogue.
Due to the small number of photons per phase bin, we did not consider performing a phase-resolved spectral analysis.

We modelled the spectrum of 1FGL J2241.9--5236 with a power law with a simple exponential cutoff shape given by
\begin{equation}
\frac{dN}{dE} = K E^{-\Gamma} {\rm exp}\left[-\left(\frac{E}{E_{\rm cutoff}}\right)\right],
\end{equation}  
where three parameters, the differential flux $K$, the photon index $\Gamma$, and the cutoff energy $E_{\rm cutoff}$ were allowed to be free in the fit. 
We found the best fit value of ($1.62 \pm 0.11_{\rm stat} \pm 0.03_{\rm sys}$) for the photon index, a cutoff value of ($4.0 \pm 0.9_{\rm stat}+0.9{\rm sys} -0.6_{\rm sys}$)  GeV, and a differential flux of K=($7.7 \pm 0.7_{\rm stat} \pm 0.5_{\rm sys}$) $\times 10^{-12}$ MeV$^{-1}$ cm$^{-2}$s$^{-1}$. The integrated photon flux above 100 MeV was found as F$_{100}$ = ($4.1 \pm 0.4_{\rm stat} \pm 0.1_{\rm sys}$) $\times 10^{-8}$ cm$^{-2}$ s$^{-1}$.  The errors quoted are the nominal $1\sigma$ statistical (first value) and systematic (second value) errors. The uncertainty in the LAT effective area is estimated to be 5\% near 1 GeV, 10\% below 0.1 GeV and 20\% over 10 GeV. The resulting systematic errors on the three spectral parameters, propagated from the uncertainties on the LAT effective area, were calculated using a set of modified IRFs bracketing the nominal (Pass6 v3) one. The best fit spectrum in this full energy band is plotted as the solid curve in Figure 6. Only statistical errors were shown and taken into account in calculating the butterfly curve.

%We found the best fit value of $(1.62 \pm 0.11)$ for the photon index, a cutoff value of $(3.99 \pm 0.90)$\,GeV, and a differential flux of $K =(7.67 \pm 0.71) \times 10^{-12}$\,MeV$^{-1}$\,cm$^{-2}$\,s$^{-1}$. The integrated photon flux above 100\,MeV was  found as F$_{100}=(4.14 \pm 0.43)\times 10^{-8}$\,cm$^{-2}$\,s$^{-1}$. The best-fit spectrum in this full energy band is plotted as the solid curve in Figure~\ref{Fig:J2241Spect}. 

The flux points in the figure were obtained from independent likelihood fits in each energy bin. For those fits, we assumed a power-law spectrum with index fixed at 2  for all the point sources and diffuse emission components. In this way, we obtained the flux values for all the sources in those energy bins, assuming that the widths of the energy bins were sufficiently small that all spectral shapes can be approximated with a power law in that range. 
The highest energy bin is determined by the highest energy photon from the source with a direction within the 68\% containment radius defined by the instrument point-spread function.
For the energy bins where the source was not detected with TS $>9$, we calculated an upper limit for the flux at 95\% confidence level.
In order to test the assumption of a cutoff shape in the spectrum of 1FGL J2241.9--5236, we also fitted its spectrum with a power-law model.
As expected for pulsar emission, this model was rejected by $6.8\sigma$ using the likelihood ratio test~\citep{mbc+96}.

\begin{figure}
\centering
\includegraphics[width=1.0\columnwidth,trim=0 5 0 10]{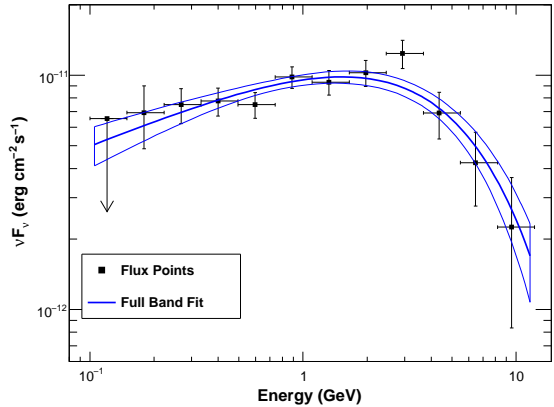}
\caption{The gamma-ray spectrum of 1FGL J2241.9--5236, the gamma-ray source associated with PSR J2241--5236. The butterfly curve (solid lines) show the best-fit spectral shape of power law with exponential cutoff in the full energy range of $0.1-300$\,GeV. The flux points were obtained from independent fits in each energy bin as explained in the text. }
\label{Fig:J2241Spect}
\end{figure}

\section{Conclusion}
Observations of six gamma-ray pulsars discovered by the {\it Fermi} satellite were carried out using the Parkes radio telescope.
We did not detect radio pulsations above a flux density limit of $\sim 0.04$~mJy.
Implications of these results are discussed in detail in \citeN {fermi_8bsp} and \citeN{rkp+10}.

In addition we have carried out deep radio observations of eleven unassociated gamma-ray sources.
Periodicity searches of these data have yielded the discovery of the
millisecond pulsars J1103--5403 and J2241--5236 and the long period
pulsar J1604--44, and the detection of the MSP J1514--49.
For the seven others, no radio pulsations were detected above a phase-averaged flux density limit of $\sim 0.04$~mJy.
This limit assumes a pulse duty cycle of $10\%$, no binary motion and that the pulsar is located at the centre of the beam.

PSR J1604--44 is a weak, long period pulsar and we consider its association with 1FGL J1604.7--4443 unlikely.
Since its selection for observation, independent evidence has shown that 1FGL J1103.9--5355 is associated with the flat spectrum radio source PKS 1101--536 \cite{msm+10}.
Furthermore, timing of the millisecond pulsar PSR J1103--5403 yields a position some 9\arcmin\ from the centroid of the gamma-ray source.
Spectral and temporal analysis suggests that the gamma-ray source is indeed powered by an AGN.
This demonstrates that with sufficiently deep observations it is possible to detect even rare objects by chance.

PSR J2241--5236 has a spin period of 2.2~ms and orbits a very low mass companion with a 3.5 hour orbital period.
The relatively high flux density and low DM of PSR J2241--5236 makes it an excellent candidate for high precision timing experiments.
When the photons from 1FGL J2241.9--5236 are folded with the radio timing ephemeris, a complex gamma-ray profile is observed.
The association of PSR J2241--5236 with 1FGL J2241.9--5236 is therefore clear.
This is consistent with the gamma-ray spectral properties and the association with an X-ray source.

\section{Acknowledgements}
The Parkes Observatory is part of the Australia Telescope which is funded by the Commonwealth of Australia for operation as a National Facility managed by CSIRO.

The \textit{Fermi} LAT Collaboration acknowledges generous ongoing support
from a number of agencies and institutes that have supported both the
development and the operation of the LAT as well as scientific data analysis.
These include the National Aeronautics and Space Administration and the
Department of Energy in the United States, the Commissariat \`a l'Energie Atomique
and the Centre National de la Recherche Scientifique / Institut National de Physique
Nucl\'eaire et de Physique des Particules in France, the Agenzia Spaziale Italiana
and the Istituto Nazionale di Fisica Nucleare in Italy, the Ministry of Education,
Culture, Sports, Science and Technology (MEXT), High Energy Accelerator Research
Organization (KEK) and Japan Aerospace Exploration Agency (JAXA) in Japan, and
the K.~A.~Wallenberg Foundation, the Swedish Research Council and the
Swedish National Space Board in Sweden.

Additional support for science analysis during the operations phase is gratefully
acknowledged from the Istituto Nazionale di Astrofisica in Italy and the Centre National d'\'Etudes Spatiales in France.

We would like to thank the University of Swinburne Centre for Astrophysics and Supercomputing for providing use of their supercomputer facilities to undertake the radio searches discussed in this paper.

\bibliographystyle{mnras}
\bibliography{journals,myrefs,modrefs,psrrefs,crossrefs}

\end{document}